\documentclass[10pt]{article}

\usepackage{mathrsfs}
\usepackage{epic, eepic}
\usepackage{subfigure}
\usepackage{amsfonts, amssymb, amsmath, graphicx, epsfig, lscape, capt-of}
\usepackage{url}
\usepackage{ifpdf}
\usepackage[boxed]{algorithm2e}
\usepackage{array}
\usepackage{booktabs}
\usepackage{multirow}
\usepackage{xcolor}
\usepackage{adjustbox}

\def\inst#1{$^{#1}$}

\makeatletter

\begin{document}%


\title{Tsallis entropy for cross-shareholding network configurations}

\author{%
Roy Cerqueti\inst{1,2}\and
Giulia Rotundo\inst{3} \and
Marcel Ausloos\inst{4,5,6} }
\date{}

\maketitle

\begin{center}
{\footnotesize
\inst{1} Department of Social and Economic Sciences, Sapienza University of Rome, p.le A. Moro 5, 00185 Roma, IT\\
\texttt{roy.cerqueti@uniroma1.it}\\
\vspace{0.3cm} \inst{2} School of Business,
London South Bank University,
London SE1 0AA, UK\\
\texttt{}\\
\vspace{0.3cm} \inst{3} Department of Statistical Sciences, Sapienza University of Rome, p.le A. Moro 5, 00185 Roma, IT\\
\texttt{giulia.rotundo@uniroma1.it }\\
\vspace{0.3cm} \inst{4} School of Business, College of Social Sciences, Arts, and Humanities, Brookfield, University of Leicester, Leicester, LE2 1RQ, United Kingdom\\
\texttt{ma683@le.ac.uk}\\
\vspace{0.3cm} \inst{5} Group of Researchers for Applications of Physics in Economy and Sociology (GRAPES), Rue de la belle jardinière, 483, Sart Tilman, B-4031 Angleur, Liege, Belgium\\
\texttt{ marcel.ausloo@ulg.ac.be}\\
\vspace{0.3cm} \inst{6} Department of Statistics and Econometrics,
Bucharest University of Economic Studies,
Bucharest, Romania
\texttt{marcel.ausloos@ase.ro}\\}

\end{center}

\newpage
\begin{abstract}
In this work, we develop the Tsallis entropy  approach for examining the cross-shareholding network 	of companies traded on the Italian stock market. In such a network, the nodes represent the companies, and the links represent the ownership.
	Within this context, we introduce the out-degree of the nodes -- which represents the diversification -- and the in-degree of them -- capturing the integration. Diversification and integration allow a clear description of the industrial structure formed by the considered companies. The stochastic dependence of diversification and integration is modelled through copulas. We argue that copulas are well suited for modelling the joint distribution.
The analysis of the stochastic dependence between integration and diversification by means of the Tsallis entropy gives a crucial information on the reaction of the market structure to the external shocks, -  on the basis of some relevant cases of dependence between the considered variables.
In this respect, the considered entropy framework provides insights on the relationship between in-degree and out-degree dependence structure and market  polarisation or fairness. Moreover, the interpretation of the results in the light of the Tsallis entropy parameter gives relevant suggestions for policymakers who aim at shaping the industrial context for having high polarisation or fair joint distribution of diversification and integration. 
Furthermore, a discussion of possible parametrisations of the in-degree and out-degree marginal distribution, -- by means of power laws or exponential functions, --  is also carried out. An empirical experiment on a large dataset of Italian companies validates the theoretical framework.
\end{abstract}



\newpage

\section{Introduction}

The presence of interconnections among companies is the ground for the propagation of   shocks over the entire industrial structure of a country; see e.g. \cite{contreras} and \cite{ohn}. This evidence has led to  a  growing number of studies exploring such structure through networks theories; see e.g. \cite{luo} and \cite{pone1}.

In this respect, a single company can be intuitively seen as a network node. The ownership relationship can be represented through a network: there is a (directed) link from a company $i$ to a company $j$ if $i$ holds shares of $j$. 
For what concerns the mutual connections among companies, several contexts can be explored on the basis of the topic under investigation. We here mention connections driven by technological transfer \cite{ferraro2017}, the presence of personal relationships
\cite{Gulati, ferraro2015, ceptu2}, the interlock of directorates \cite{Bellenzier,Croci,GRAMDAASS},  and capabilities at organisational level
\cite{ceptu1,ceptu2}. For a survey on this field, see e.g. \cite{weber}. 

We propose a specific focus on the cross-shareholding matrix, which is associated to the directed links, thereby capturing the so-called in-degree and out-degree of  each node.

Specifically, the in-degree of a company -- say, $k_{in}$ -- is the number of companies
holding some ownerships of the considered node. Such a concept has a clear interpretation on the integration of any given company in  its reference industrial and productive environment.
Similarly,  the out-degree of a company -- namely, $k_{out}$ -- counts the companies included in the portfolio of the considered node. Thus, $k_{out}$ is associated to diversification, which in turn may point out to information on the possible reaction of a considered company to markets fluctuations. For the concepts of integration and diversification, we refer the interested reader to \cite{SoumaBook}. 

Notice that the so  followed approach is grounded on the existence of a connection -- in terms of ownership relations -- between two companies. In so doing, we explore diversification and integration -- along with market concentration, which is a synthesis of them -- as a matter of pure shareholding strategies and through the singular  attitude of companies to collect shares of other companies, and at the same time to have shares   own by other companies, - "other companies" which can be the same being owner and owned\footnote{Renault SA, which is part-owned by the French state, owns 43\% of Nissan Motor Co, while the Japanese firm has 15\% of the French carmaker, - but with no voting rights in this case.}. Within such a thinking, the amount of inter company flows leads to a discussion on the size of the connections between companies. In this setting, in-degrees and out-degrees should be reasonably written as sums of percentages of in-flows and out-flows. Thus, the in-degree can be high in both cases, i.e., when there is a large number of existing in-connections with small flows or small values of in-connections with large entities of flows;  the same consideration applies also for the out-degree, of course. The numerical dimension of the connections is then lost -- even if a new information on the size of the flows is available. Yet, the analysis of  such flows is clearly beyond the scopes of the present paper.

While out-degrees are widely explored, for their natural connections with the resilience of an industrial system, see e.g. \cite{Newman, soromaki,gao, iori, delpini}, scarce attention has been paid to in-degrees.  
Let us point to a noticeable contribution on the trade-off between diversification and integration in the analysis of economic crises   in \cite{Elliott}. 

Here, we are concerned by the market concentration, - which represents a synthesis of diversification and integration, by means of the entropy of the in-degree and out-degree distributions. The entropy concepts allows to understand the position of the considered industrial structure between the extreme cases of uniform diversification and integration and {\it a contrario}  strong polarisation, with only one company playing the role of the leader.

Furthermore, we include   also a deep analysis of the particular features of the distributions through a generalised concept   \cite{Tsallis,Masz} of  Boltzmann-Gibbs  (or equivalently  Shannon information \cite{Shannon,Masz,EPJB86.13GRMAcomplex}) entropy. To this end, we move from \cite{Entropy1} and deal with the Tsallis entropy  for discussing the in- and out-degrees   distributions of the companies. 

Tsallis entropy -- introduced in \cite{Tsallis} -- has been applied in a number of contexts related to economics and finance; see e.g. the excellent review in \cite{Tsallis1} and references therein. Most of the time, the   studies  concern risk or portfolio management \cite{Devi19, mbraz, Vinte,MaasoumiRacine02,AusloosIvanova03dynamical,zhou13applications}.
Our present report   seems to be the first contribution dealing with this powerful instrument in the context of the cross-shareholding matrix for  its related network of companies. 

Tsallis entropy depends on a (usually real, see a complex case in  \cite{complexq}) parameter, whose interpretation provides a relevant information on the shape of the distributions. Indeed, when the parameter is negative (positive), then Tsallis entropy 
attains its maximum in the highest polarisation case (in the uniform distribution case). Moreover, a negative value of the parameter is associated to a strong relevance of   fat tails and rare events; see e.g.  \cite{Masz}.

To explore in depth the relationship between integration and diversification, we propose the analysis of the joint distributions between such terms in the relevant cases of independence -- i.e., when the stochastic dependence is described by a product copula -- and in the maximum level case of positive (negative) dependence -- i.e., when the dependence is given by the upper (lower) Frechet bounds copulas. These represent the mathematical bounds of the set of the copulas corresponding to the cases of perfect positive (negative) correlation; see \cite{frechet}.  
For a complete description of the concept of copulas and on how it serves as modelling stochastic dependence, see e.g. \cite{joe,nelsen} and refer to the Sklar's Theorem \cite{sklar}. Indeed, Sklar's Theorem provides a reading of the copulas as mathematical functions transforming the marginal distributions of a set of random variables into their joint distribution (see also below). 

To validate our theoretical proposal, we consider a high-quality dataset of holdings listed in the Italian Stock Market.  Such a selection,  the Italian Stock Market as reference context, has been driven by data availability. Indeed, the phase of data collection has been particularly challenging, with manual collection procedures and matching among different datasets -- see the details in Section \ref{sec:data}. Of course, the premise of the data collection procedure is data availability. This said, even if it is theoretically easy to reproduce the analysis for all the major markets, -- like the US and the UK ones,   the practical implementation in different contexts requires a non-trivial effort and data availability.

We also propose an extension of the analysis to a wide and universal economic system, where in-degrees are assumed to be synthesised by two parametric functions of either power law or exponential types, while  the out-degree distribution obeys a power law; see e.g. \cite{Caldarelli}. In particular, we have included the parameters of such functions in the calibrating quantities set. 
Such a proposed extension leads to useful discussions about the assessment of missing links in the cross-shareholding matrix, in line with some literature contributions, like e.g. \cite{Axel,Cimini,Serri}.

Some  results emerge from our study. The obtained outcomes suggest strategies that should be implemented by  policymakers if pursuing a highly polarised industrial structure goal, -- with a company holding the shares of all the other ones and, at the same time, whose shares are included in the portfolios of the others, -- or a fair joint distribution of diversification and integration. Such policies are built on the basis of the dependence structure between in-degrees and out-degrees and on enforcing the shapes of their distributions in a proper way.

The remaining part of the paper is organised as follows. Section \ref{revlit} provides some information on the reference literature on cross-shareholding. Section \ref{sec:method} gives the details of the methodological devices used in the analysis.  Section \ref{sec:data} provides a description of the dataset employed for the methodological validation, and in particular the network construction in Subsection  \ref{sec:construction}.  Section \ref{sec:results} describes and discusses the obtained findings.    Conclusions and comments on policy implications are found in  Section  \ref{sec:conclusionpolicy}.

\section{Brief review of the reference literature on cross-shareholdings}
\label{revlit}

This section provides a list of key papers dealing with cross-shareholdings. Such a list is not exhaustive, but the referred contributions are particularly close to the present study -- even if they present remarkable differences.
As a premise, we have to state that the framework adopted in this paper is quite new compared to other papers on the cross-shareholdings.
 
In \cite{Derrico},  a complex networks approach is used for identifying the companies  which are central in the information flow and for the control. The coupling among in-degree and out-degree is not examined explicitly, although it intervenes in the empirical estimates of the flow-betweenness and of  other centrality measures.
 
The perspective in Abreu et al. \cite{Abreu} is of an empirical nature, without a precise focus on the relationship between in-degrees and out-degrees,   i.e., as integration and diversification measures, respectively.
 
In \cite{Bebchuk}, the possibility to use cross-shareholdings for achieving the control of companies through intermediaries is examined, but there is again no focus on the relationship between integration and diversification as  optimal means toward the considered specific targets.

Vitali et al. \cite{Vitali} offer the analysis of the structure and topology of the transnational ownership network of cross-shareholdings. This is a pretty empirical paper, without further steps in the analysis of the stochastic dependence on integration and diversification.

An analysis of the relevance of the cross-shareholdings in the Japanese markets can be found in \cite{Lichtenberg}. 
The target of the quoted paper is to understand the role of shareholdings in order to reduce the risk/performance ratio. However, the focus is quite different from the one tackled in this present paper.
 
In \cite{Okabe}, Okabe performs an economic analysis on cross-shareholdings in Japan, where this theme is quite relevant. Trends and implications for the Japanese economic system and related public policies are discussed. However, the analysis is mostly performed from the perspective of economics rather than by proposing novel methods for the investigation.

The framework of the stochastic dependence among integration and diversification considered in the present paper is close to that in \cite{GRAMDQuQu}, but presently under a wider viewpoint; in \cite{GRAMDQuQu}, one  uses a rewiring procedure as methodological instrument.

\section{Methodology}
\label{sec:method}

This section describes the techniques and the tools used for achieving the targets of the analysis.
%

\subsection{Preliminaries and notations} 
First, we   introduce the main concepts that are used in the paper.

Given a node $j \in V$, the \emph{in-degree} $k_{in}(j)$ represents the integration, i.e. the number of companies owning shares of company $j$. It is defined as follows:
$$
k_{in}(j)=\sum_{i=1}^N a_{ij}
$$ 
In the same line, given $i \in V$, the \emph{out-degree} $k_{out}(i)$ represents the diversification, i.e. the number of companies in the portfolio of company $i$. It is defined as follows:
$$
k_{out}(i)=\sum_{j=1}^N a_{ij}.
$$
Both $k_{in}$ and $k_{out}$ have to be considered here as random variables, whose empirical distributions are obtained by considering the real data described in Subsection \ref{sec:data}.

The cumulative distribution functions of $k_{in}$ and $k_{out}$ is denoted by $F_{k_{in}}:\mathbb{R}\to [0,1]$ and $F_{k_{out}}:\mathbb{R} \to [0,1]$, respectively. Their joint distribution is denoted by $F_{k_{in},k_{out}} :\mathbb{R}^2\to [0,1]$.

The generic joint distribution function $F_{k_{in},k_{out}}$ is associated to a bivariate density function. It is discrete, in the empirical case we are treating; the distribution is denoted by $\mathbf{p}=(p_{ij}: i=1, \dots, n; j=1, \dots, m)$ such that
\begin{equation}
	\label{pij} p_{ij}=Prob(k_{in}=i,k_{out}=j), \qquad \forall \, i, \,j,
	\end{equation}
with $$
\sum_{i,j}p_{ij}=1.
$$
The values of the integers $n$ and $m$ will be properly fixed in the subsequent empirical analysis. 

In the sequel, for such a bivariate probability distribution,   we compute the Tsallis entropy, usually defined as  
\begin{equation}
	\label{tsal} S_q=\frac{1}{q-1} \left(1-\sum_{i,j}p_{ij}^q\right),
	\end{equation}
where $q \in \mathbb{R}$ is the Tsallis parameter.

A bivariate copula $C:[0,1]^2 \to [0,1]$ (see e.g. \cite{nelsen})  is a special function able to describe the dependence structure between two random variables through the classical Sklar's Theorem (see \cite{sklar}). We enunciate such a crucial result by employing the notation used in the present paper.

\vspace{0.3cm}

 {\bf Sklar's Theorem reads:}

	there exists a copula $C:[0,1]^2 \to [0,1]$ such that, for each $(s,h) \in
	{\mathbb R}^2$, one has
	\begin{equation}
	\label{eq:Sklar} F_{k_{in},k_{out}}(s,h)=C(F_{k_{in}}(s), F_{k_{out}}(h)).
	\end{equation}
      If $F_{k_{in}}, F_{k_{out}}$ are continuous, then $C$ satisfying (\ref{eq:Sklar}) is
	unique. Conversely, if $C$ is a copula and $F_{k_{in}}, F_{k_{out}}$
	are distribution functions, then $F_{k_{in},k_{out}}$ in
	(\ref{eq:Sklar}) is a bidimensional joint distribution function with
	marginal distribution functions $F_{k_{in},k_{out}}$.

\vspace{1.3cm}

According to this Theorem 
copulas describe different types of stochastic dependence which could be  found between two random variables. In so doing, one is also capable to provide insights on the nature of the stochastic dependence of tis empirical joint distribution. 

We denote by $F^C_{k_{in},k_{out}}:[0,1]^2 \to [0,1]$ the joint distribution function resulting from the application of Sklar's Theorem with a generic copula $C$, according to the previous Formula (\ref{eq:Sklar}).



\subsubsection{Reasoning behind the Tsallis entropy}

This section is devoted to the justification of the selection of Tsallis entropy as a key methodological measurement device. We provide a comparison between Tsallis entropy and the well-known and largely used Gibbs entropy. 
In fact, Tsallis entropy is known  to exhibit substantial strengths when compared to the Gibbs one. To support this statement, we proceed under both technical and applied perspectives. 

From a purely mathematical point of view, Tsallis entropy represents a generalisation of the Gibbs entropy. Indeed, Tsallis entropy, formally a  fractional exponential approach,  depends on an often real (but see \cite{complexq})  parameter $q$, introduced  in Eq. (\ref{tsal});
 when $q \to 1$, the  Tsallis entropy collapses to the Gibbs entropy. Hence, the Tsallis entropy is able to capture several aspects that are not covered by the Gibbs entropy, -- all those aspects related to a not unitary parameter $q$. In our context, the main results will be seen to be related to negative $q$ values. Thus, it is clear that the Gibbs entropy would not allow us to provide a deep understanding of the nature of the stochastic dependence between in-degree and out-degree distributions. 

In the context of applied science, we may recall that classical statistical mechanics of macroscopic systems in equilibrium is based on Boltzmann's
principle and Gibbs entropy. However, Boltzmann-Gibbs statistical mechanics and standard thermodynamics present
serious difficulties or anomalies for  non-equilibrium, open, non-ergodic, non-mixing, systems, and for those which exhibit
memory retention. Within a long list, we may mention systems involving long-range interactions (see e.g. \cite{Petroni, Barbosa}), non-Markovian stochastic processes, like financial markets (see e.g. \cite{Queiros, Rak2007, Rak2013, Bila, Ruseckas, Jakimowicz}), dissipative systems in a phase space which has some underlying looking (multi)fractal-like structure (see e.g. \cite{Pavlos}), like many open social systems, all hardly having an additive property (see e.g. \cite{Gell-Mann}).

In brief, Tsallis theory provides a better thermo-statistical description than the standard Boltzmann-Gibbs formalism, because the Tsallis fractional exponential approach allows to encompass cases of non-equilibrium and dissipative systems into hard core statistical mechanics principles.

\subsection{Outline of the analysis}

The analysis is carried out  in two main directions.

First, we compute and discuss the Tsallis entropy of the joint distribution $F^{C}_{k_{in},k_{out}}$, which is obtained by applying the Sklar's Theorem with some specific copulas $C$. In so doing, we provide useful insights on the behaviour of the cross-shareholding system under different scenarios of interactions between in-degrees and out-degrees.

In particular, we address the corner cases of maximal positive and negative dependence, and the case of independence. Such cases correspond to the following copulas:
\begin{itemize}
	\item  Product (independence)
      \begin{equation}
       \label{ind} C_P(u,v)=uv
      \end{equation}
	\item Lower Frechet (maximal negative dependence) and Upper Frechet (maximal positive dependence) 
	\begin{equation}
	\label{LF-UF} C_{LF}(u,v)=\max\{u+v-1,0\}, \qquad C_{UF}(u,v)=\min\{u,v\}.
	\end{equation}
\end{itemize}

Second, we discuss the sensitivity analysis of the in- and out-degrees distributions when they are properly parametrised, by means of the Tsallis entropy.

In this respect, while the literature points out the ubiquitous presence of a power law for the out-degree distribution, the in-degree is much less studied. However, the main theoretical functions which can be suitably used for approximating the  in-degree empirical distribution are either the power law  or the exponential law (see \cite{Entropy1} and references therein contained). Therefore, on one side, we consider the marginal distribution of the out-degree as following a power law; on the other side, we consider two cases, power law or exponential function for modelling the  in-degree empirical distribution.

The power law and the exponential law for a generic discrete random variable $X$ are defined as follows:

\begin{itemize} 
\item Power law:  
\begin{equation}
\label{power}
Prob(X=x) = ax^{-k},
\end{equation}
where $x \geq 0$, $a>0$ is a normalising constant and $k>0$.
\item Exponential law:  \begin{equation}
\label{exp}
Prob(X=x) = ae^{-k x},
\end{equation}
where $x \geq 0$, $a>0$ is a normalising constant and $k>0$.
\end{itemize}

Thereafter, we implement the sensitivity analysis in three cases:
 
\begin{itemize}
	\item[A)] under the hypothesis of $k_{out}$ described by a power law as in (\ref{power}) and $k_{in}$ has its empirical distribution, the power law exponent $k$ is allowed to change and is treated as a parameter;
	\item[B)] under the hypothesis of $k_{in}$ power law as in (\ref{power}) and $k_{out}$ empirical: the power law exponent $k$ is allowed to change and is treated as a parameter;
	\item[C)] under the hypothesis of $k_{in}$ exponential as in (\ref{exp}) and $k_{out}$ empirical: the parameter $k$ in the exponential is allowed to change, as any parameter does.
\end{itemize} 

Thus, in each case,  there are 2 parameters: $q$ for the Tsallis entropy, $k$ for the power law or exponential.
In all the cases, we have employed the three copulas $C_{I}$, $C_{LF}$ and $C_{UF}$ introduced in (\ref{ind}) and (\ref{LF-UF}) for deriving the joint probability distribution, according to  Sklar Theorem.

%


\section{The network}
\label{sec:network}
We here present the cross-shareholding network that are used in the analysis.

\subsection{The data}   
\label{sec:data}
We consider the data already used in \cite{GRAMD,Entropy1}. The dataset gathers data of the Milan Stock Exchange (MIB30) on May 10th, 2008. First, data were obtained through the CONSOB database. For each company $j$, an informative page is shown, which contains the information on the holdings, that is the list of companies $i$, traded in the same market, which the shares of $j$ are sold to. The set of all the couples $(i,j)$ constitutes the matrix of cross-holdings. 
  CONSOB is the major surveillance body for the Italian Stock Market. CONSOB verifies the transparency of market operations; it has the power to stop the market in case of excess of losses/returns,; CONSOB controls the proper disclosure of information. Unfortunately,  CONSOB records only the holdings above 2\%.
 Therefore, the data was  cross-checked through the Bureau Van Dijk platform.
 
Differently from the database of prices of the shares, there is no command which allows to download all the data at once. The data gathering requires manual opening of each file, and manual storing of the relevant information. Moreover, the way in which the companies are named is not uniform: sometimes shortcuts are used instead of the original extended names. Therefore, the data collection cannot be done automatically "blind folded". The data has also to be gathered at a selected date: it is like taking a picture of the actual situation of the market on a specific day. The time needed for gathering  the data and finalising  the sample is quite long, since the data was manually cross-checked with other databases. Notice that the data on  banks was   cross-checked with the BANKSCOPE database, which, as the name suggests, is specifically focusing on banks,  whence not reporting data on   other companies.  

On the other side, AIDA provides  some complementary information, since  AIDA contains  information on all  companies, - apart from banks. The cross-checking was necessary to be sure that we include in the database all ownerships due to investments and all cross-relationships among companies, - yet excluding some very minor ones due to the management of portfolios by mutual funds. Alas, some companies had very incomplete data. Finally, the resulting sample contains the cross-holdings of 247 stocks of companies. They represent   94\% of the total amount of MTA segment (MTA stands for Borsa Italiana's Main Market, that is Italian Main Stock Market. MTA is a regulated market subject to stringent requirements in line with the expectations of professional and private investors.). The sample corresponds to  95.22\% of the total capitalisation on that date,  May 10th, 2008, which nevertheless makes the analysis quite suitable for a whole outlook about the links among the most relevant traded companies. Notice that the total number of cross-ownership is 243, thus less than the number of   companies. In fact, there are companies traded in the Italian Stock Market, which do not buy or sell shares of other companies traded in the Italian market. 

The vast majority of holdings is due either to industrial purposes, or to an internal organisation of companies: for instance, the energy company ENI owns shares of two other companies, SAIPEM  and  SNAM RETE GAS, with the a specific focus on gas delivery management. Another example is given by the financial company IFIL which is managing the financial parts of FIAT (now merged in FCA) and JUVENTUS (football club). In turn, IFI PRIV owns the "privileged" part of IFI,  belonging to the Agnelli  family.

The number of companies holding shares of $k$ other companies decreases sharply as $k$ increases. In fact, there are 72 companies owning   shares of only 1 other company; 16 companies owning shares of 2 other companies; only 7 and 6 companies are owning shares of 3 and 4 other companies, respectively. There are only 0 or 1 companies holding shares of 6 or more other companies;  the maximum ownership in  19  companies is due to the insurance company "Assicurazioni Generali", that uses  ownership as part of its institutional mission. 
  
A symmetric question holds: which is the number $h$ of companies  to which a specific company has sold  shares?  According to the literature on this topic, the question is less popular than the previous one. In our specific dataset, the maximum value of $h$ is 10; 
there are 84 companies which sell their shares to  only 1 company;   29  companies   sell their shares to two companies;  15 are  selling to three; only    5  companies have sold to four other companies, and another 5 are selling to more than four companies. Therefore, roughly speaking, the very prevailing behaviour is the relation through a sale of shares to only one other company in the market.

\subsection{Construction of the network}
\label{sec:construction}

The firms are represented by the nodes of an unweighted network. We collect them in a set $V= \{1,\cdots, N\}$. If a company $j$ is held by company $i$, then there is a directed link from $i$ to $j$. The links are collected in a set $E$. In so doing, we create a network $(V,E)$, whose adjacency matrix $A=(a_{ij})_{i,j \in V}$ is a $N\times N$ matrix such that $a_{ij}=1$ if $(i,j)\in E$ and $a_{ij}=0$ otherwise.

The insulated nodes have been removed from the analysis; the giant component and the small connected components being kept,  the network is made of 158 nodes, whence the adjacency matrix is 158 x 158. 

 \section{Results and discussion}
 \label{sec:results}
 
We report here the results of the analysis, along with a discussion of these.

As a premise, we set $n=10$ and $m=19$, in accord to the maximum values of $k_{in}$ and $k_{out}$ which are observed in the empirical dataset.

It is immediate to observe that  the Tsallis entropy $S_q$ in Eq. (\ref{tsal}) is strictly decreasing with respect to the parameter $q \in \mathbb{R}$, with an  asymptotic behaviour given by
$$
\lim_{q \to -\infty}S_q=+\infty; \qquad \lim_{q \to +\infty}S_q=0.
$$

This said, we restrict our graphical representations of the behaviour of the Tsallis entropy with respect to $q$ to a small interval including zero, for a better visualisation of the outcomes.

\begin{figure}
  		\centering
  		\includegraphics[width=0.7\linewidth]{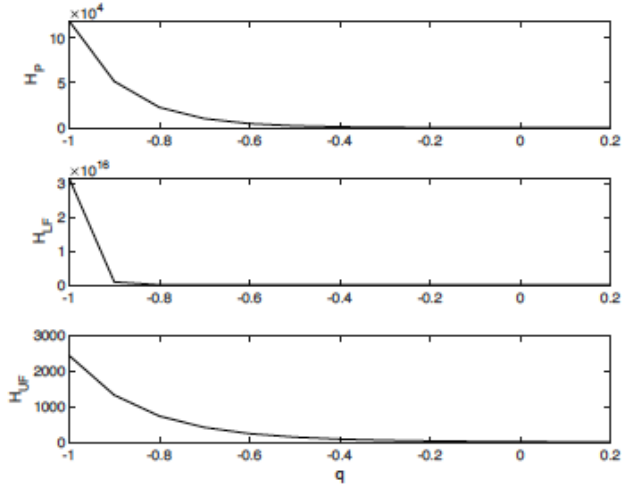}
  		\caption{The Tsallis entropy $H=H_P, H_{LF}, H_{UF}$ as a function of $q$,  in the cases of copula $C=C_P, C_{LF}, C_{UF}$ as in (\ref{ind}) and (\ref{LF-UF}) -- upper, middle and lower panel, respectively.} 
  		\label{fig:TsallisNonParametric}
  	\end{figure}

Figure \ref{fig:TsallisNonParametric} shows the behaviour of the values of the Tsallis entropy as the parameter $q$ varies, in the three cases of joint distributions, $F^{C}_{k_{in},k_{out}}$ with $C=C_P, C_{LF}, C_{UF}$ as in (\ref{ind}) and (\ref{LF-UF}), -- in upper, middle and lower panel, respectively.

The Upper Frechet bound is the one with the slowest decrease; it is substantially flat with respect to the other cases. Moreover, the   Lower Frechet bound is associated to very high values of the Tsallis entropy when $q $ approaches -1; such a case is also the one presenting a very rapid collapse of   $S_q$ as $q$ increases. 

An interpretation of these results is in order. The predominance -- to be intended as the highest values of Tsallis entropy -- of the case of copula $C_{LF}$ means that the joint probability between in-degree and out-degree is highly polarised when there is a perfectly negative correlation between such quantities. This is particularly true when $q$ is negative; hence, the fat tails of the distribution do play a key role in determining such an outcome. Results change when moving to the independence and to the maximum level of positive dependence. In particular,   the Upper Frechet case corresponds to the highest similarity between  the uniform case and  the considered joint probability distribution. 

The policymaker should then force the in-degrees and out-degrees of the companies to exhibit similar patterns -- i.e., integration and diversification should coincide, -- when the target is  a homogeneous industrial structure; {\it a contrario},  integration and diversification should be forced to exhibit a large discrepancy, if the aim of the policymaker is to foster the predominance of a company over the others.

We now deal with the cases A), B) and C) described in the previous section, which are related to different parametrizations of the in- and out-degree marginal distributions. 

\begin{figure}
	\centering
	\includegraphics[width=1.1\linewidth]{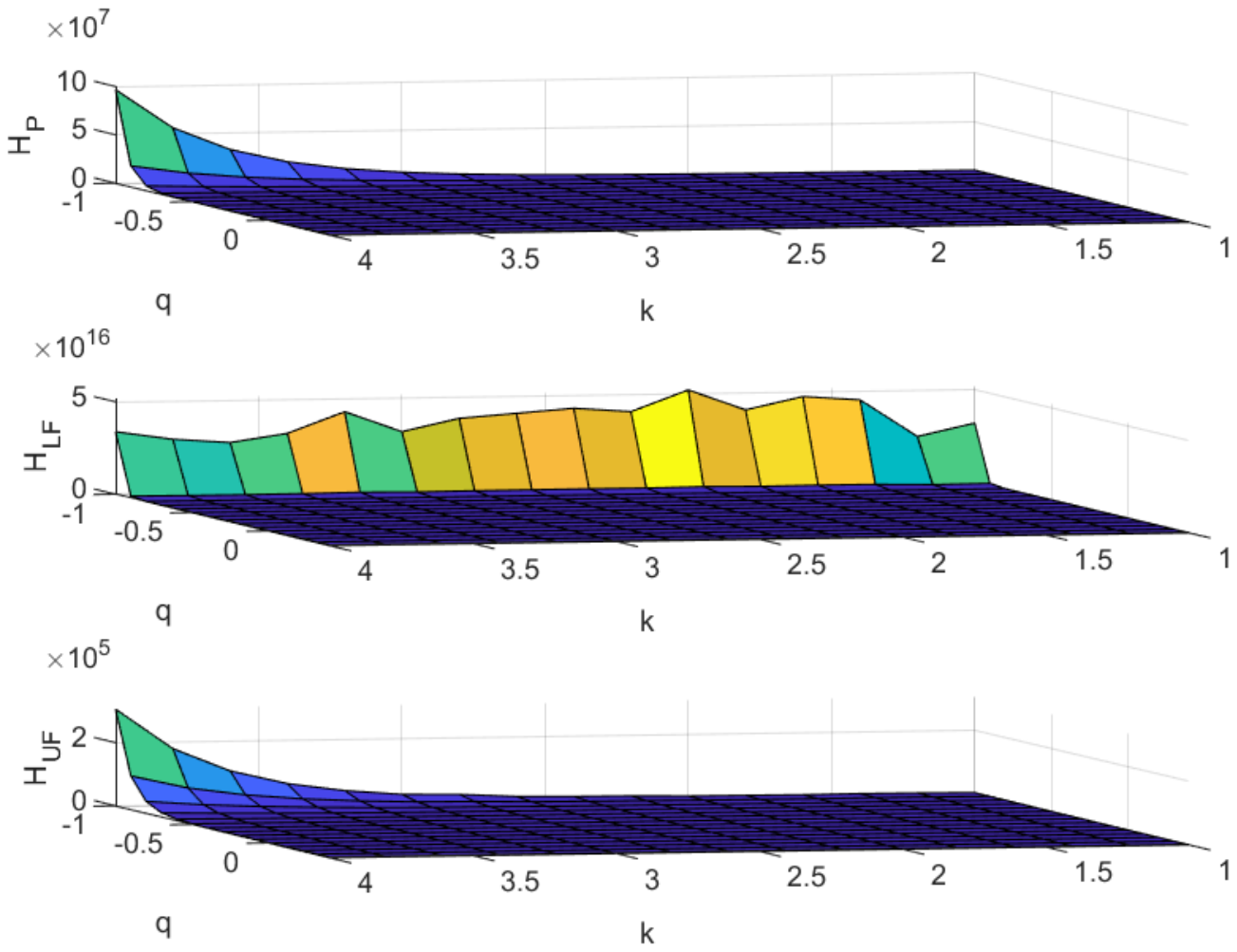}
	\caption{Tsallis entropy as a function of its parameter $q$ and the exponent of the power law $k$ for the out-degree. All the cases of copula $C=C_P, C_{LF}, C_{UF}$ as in (\ref{ind}) and (\ref{LF-UF}) -- upper, middle and lower panel, respectively -- are reported.}
	\label{fig:KoutPowerNonParam}
\end{figure}

\begin{itemize}
\item[A)] $k_{out}$ is described by a power law as in (\ref{power}), while $k_{in}$ is taken with its empirical distribution.
\end{itemize}
Figure	\ref{fig:KoutPowerNonParam} shows the Tsallis entropy as function of its parameter $q$ and the exponent of the power law $k$ for the cases of copula $C=C_P, C_{LF}, C_{UF}$ as in Eq. (\ref{ind}) and Eq. (\ref{LF-UF}) -- upper, middle and lower panel, respectively. 

In all  cases, we observe that Tsallis entropy is decreasing as $k$ decreases and $q$ increases. The growth toward infinity is very rapid as $q$ approaches -1. This behaviour is more evident when $k$ assumes large values, i.e. when the probability that $k_{out}$ assumes a large value is particularly small, -- and when in-degree and out-degree are highly positively correlated or are uncorrelated. If in-degree and out-degree have the maximum level of negative correlation, then the same behaviour seems to be rather independent from the value of the power law parameter. The apparent crests on $H_{LF}$ actually correspond to very high values of  $H_{LF}$; furthermore, the case with $C_{LF}$ is confirmed to have the highest level of Tsallis entropy.   

We can read the results by stating that the joint probability of in- and out-degree shows a high level of polarisation in presence of a perfectly negative correlation. Such a finding does not depend on the specific parametrisation of the out-degree through a power law. Differently, we see polarisation only for $k$ large enough when the cases of stochastic independence or perfectly positive correlations are considered. This behaviour is amplified for negative $q$ values,  hence giving credit to the action of the fat tails of the distribution in so determining it. 

The policymaker has now two devices for shaping the considered industrial structure.  Beyond dealing with the dependence between diversification and integration, --   we refer to the comments stated above for Figure \ref{fig:TsallisNonParametric}, -- she/he can also force the individual companies to form specific out-degrees distributions. Indeed, in the particular cases of independence and maximum positive correlation, one can obtain some polarisation by shaping the out-degrees in order to obtain a low probability of having large values, -- i.e., by taking large values of the parameter $k$. Such an action is not needed when the correlation between in-degree and out-degree is of perfectly positive type.

\begin{itemize}
\item[B)] $k_{in}$ is a power law as in (\ref{power}) and $k_{out}$ has its empirical distribution
\end{itemize}

\begin{figure}      
	\centering
	\includegraphics[width=1\linewidth]{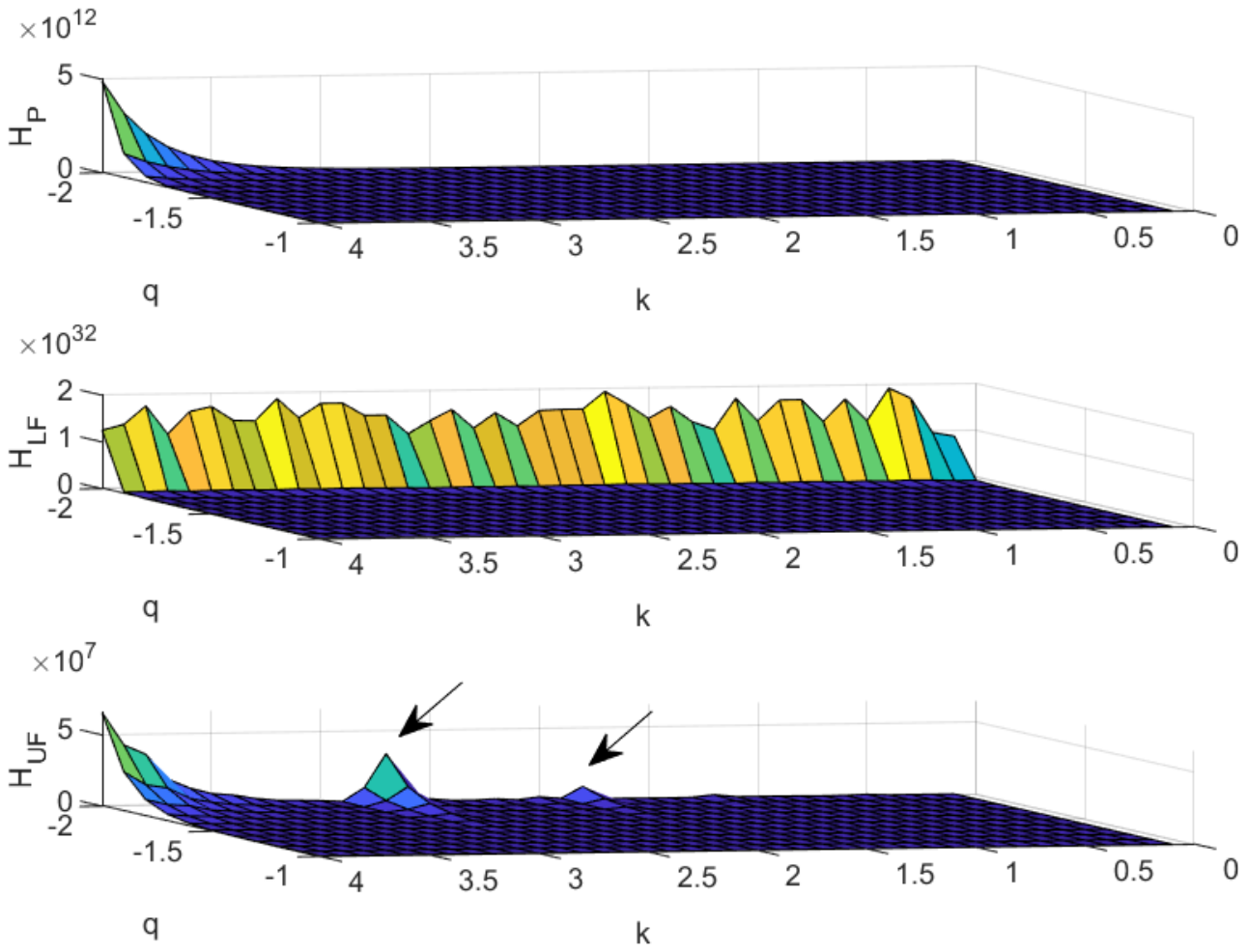}
	\caption{Tsallis entropy as function of its parameter $q$ and the exponent of the power law $k$ for the in-degree. The cases of copulas $C_P, C_{LF}, C_{UF}$ as in (\ref{ind}) and (\ref{LF-UF}) are presented in the upper, middle and lower panel, respectively.} 
	\label{fig:KinPowerNonParam}
\end{figure}

Figure	\ref{fig:KinPowerNonParam} presents the values of the Tsallis entropy as a function of $q$ and $k$. Also in this case, copulas $C_P, C_{LF}, C_{UF}$ as in (\ref{ind}) and (\ref{LF-UF}), are in the upper, middle and lower panel, respectively. For a better visualisation of the results, we  display only when $q<0$.

The behaviour of  the Tsallis entropy is quite similar to that of case A), with four noticeable exceptions. Firstly, the scales are completely different. The values of the Tsallis entropy are much higher in this case than in case A). Secondly, to appreciate the decreasing behaviour of the Tsallis entropy, one needs to take $q$ close to -2, instead of $q=-1$, as in the previous case. Thirdly, we observe a deviation in the case of perfectly negative correlation, with two lines of local maxima occurring at $q \simeq    -2$, for $k=2.7$ and $k=1.8$ (see the arrows in Figure \ref{fig:KinPowerNonParam}). Fourthly, the crest appearing in the case of perfectly negative correlation is much more jagged   than in case A). 

The similarities between cases A) and B) insure that all comments raised for A) remain valid also for this case B). 
The presence of   local maxima and  the jagged crest  do point to the questionability of the power law parameter as a device for controlling the polarisation of the joint distribution between in-degree and out-degree when   the   value of $q$ is at its minimum. This is particularly evident for the case of perfectly negative correlation, -- i.e., in the case of jagged crest, -- while an action for properly calibrating the parameter $k \simeq    -2.7$  and 
$\simeq 1.8$ remains   possible for the case of perfectly positive correlation.

\begin{itemize}
\item[C)] $k_{in}$ has an exponential distribution as in (\ref{exp}) and $k_{out}$ has its empirical distribution.
\end{itemize}

\begin{figure}
	\centering
	\includegraphics[width=1\linewidth]{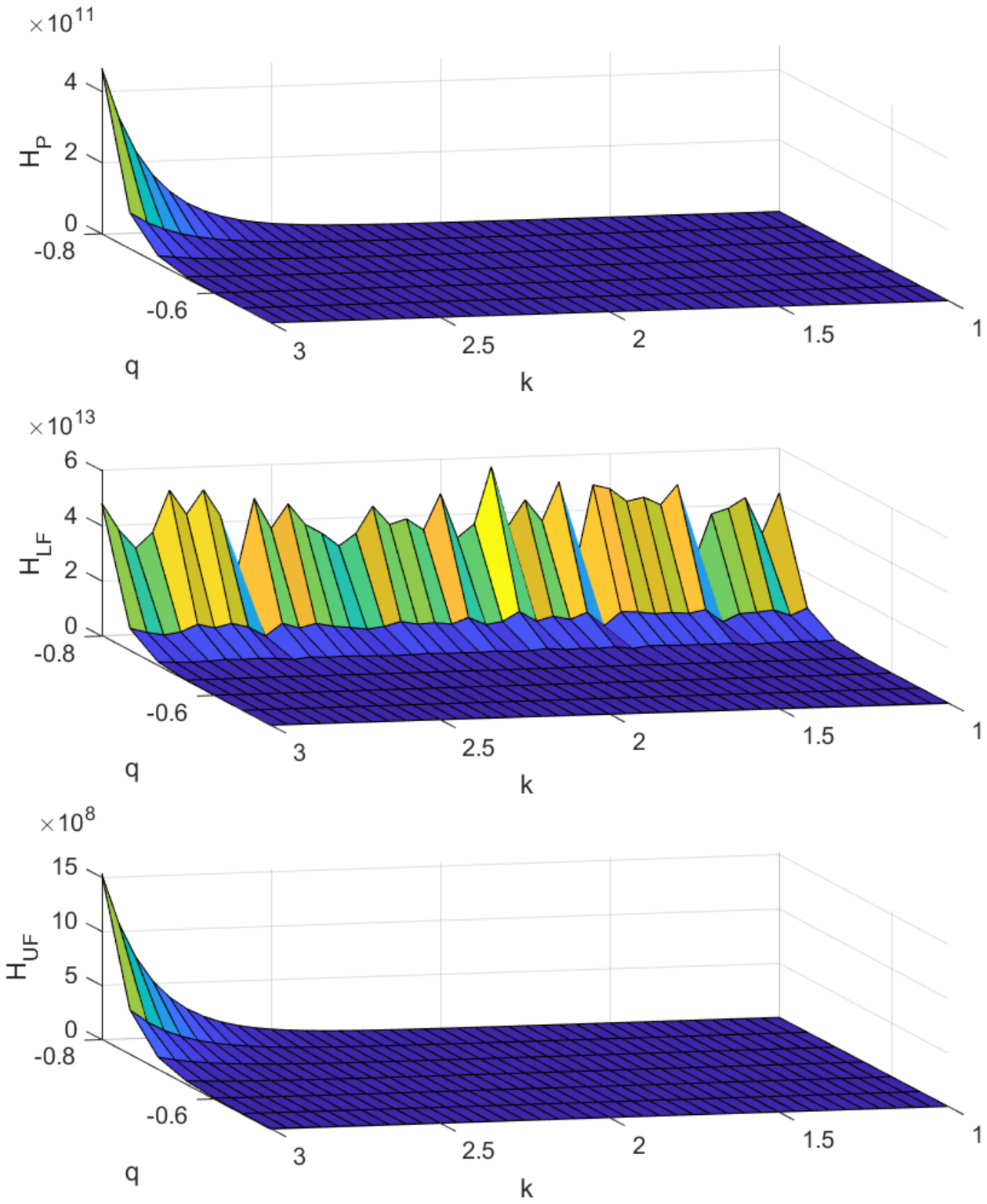}
	\caption{Tsallis entropy as a function of parameter $q$ and  $k$ for describing the exponential decrease of the in-degree. The cases of copulas $C_P, C_{LF}, C_{UF}$ as in (\ref{ind}) and (\ref{LF-UF}) are described in upper, middle and lower panel, respectively.}
	\label{fig:KinExpNonParam}
\end{figure}

The upper, middle and lower panel of Figure \ref{fig:KinExpNonParam}  display the Tsallis entropy as a function of $q$ and $k$, for copulas $C_P, C_{LF}, C_{UF}$ as in (\ref{ind}) and (\ref{LF-UF}), respectively; for a clear view of the behaviour of the surface, we  present $q<0$ only.

As for B), also  the behaviour of Tsallis entropy is analogous to the one observed for A), but with three main differences. Indeed, the decreasing behaviour of the Tsallis entropy can be properly visualised for $q$ close to -0.8 (it was -1 and -2 in cases A) and B), respectively); moreover, the crest appearing in the middle panel at low values of $q$ is more jagged here than in A); finally, the minimum value of $q$ appearing in Figure \ref{fig:KinExpNonParam} is -0.8 instead of -1 (case A)) and -2 (case B)).

Some relevant insights can be derived by comparing the three cases A), B) and C). When the desired target is to shape the cross-shareholding network for   a polarised situation, -- with a company holding the widest part of shares of the others and, at the same time, whose shares are in the portfolios of the other companies, -- then one has to impose a perfectly negative dependence between the in-degree and the out-degree.  Moreover, one has also to shape the distribution of the in-degree as a power law; this means that the probability of having a high  in-degree value has to be lower than that of having a low in-degree value.  Lastly, the joint distribution between in-degree and out-degree should include also the presence of fat tails, so that one can employ the informative content of the Tsallis entropy in the case of large negative value of $q$. Under the conditions described above, the Tsallis entropy attains its highest value -- see case B), middle panel.
Differently, by imposing the maximum level of positive dependence and a power law behaviour on the out-degree distribution, with a small value of the parameter $k$, one pursues the objective of shaping the industrial structure towards a more uniform integration and diversification;  see case A), lower panel.

 \section{Conclusions and policy implications}  \label{sec:conclusionpolicy}
 
To conclude, we can offer some  general remarks on policy implications.

The starting point of the analysis is to describe the industrial structure of a country, -- in terms of market integration and diversification and, consequently, of concentration. In this respect, the policy makers might aim at fostering the competition in the market or, conversely, at shaping the market for having a leading company. 

This theme is of paramount relevance for policy makers. Indeed, the interest of regulatory Authorities in the raise of concentration is witnessed by its explicit insertion in official documents. For instance, the study of the classical Herfindahl-Hirschman index (HHi) -- which is a relevant measure of market concentration -- plays a significant role in the assessment of possible enforcement of US antitrust laws \cite{US}. Since 1982, the Merger Guidelines by the U.S. Department of Justice and the Federal Trade Commission\footnote{https://www.justice.gov/atr/horizontal-merger-guidelines-0} have provided an indication for the identification of post merger markets as "unconcentrated", mildly concentrated, or highly concentrated based on the value of HHi. For a more scientific perspective, we refer e.g. to \cite{Crane} and \cite{GRAMDQuQu}. In this respect, we mention also \cite{Elliott}, who have shown that some peculiar combinations of integration and diversification might lead industrial structures to be more vulnerable to financial fluctuations.

\vskip1cm
 {\bf Acknowledgments}\\The authors thank Prof. Anna Maria D'Arcangelis for providing data and for many fruitful discussions on the policy implications of the analysis. 

\bibliographystyle{mdpi}

\begin{thebibliography}{999}

\bibitem{Abreu} Abreu, M. P., Grassi, R., Del-Vecchio, R. R., Structure of control in financial networks: An application to the Brazilian stock market. {\em Physica A: Statistical Mechanics and its Applications}  {\bf 2019}, 522, 302-314.

\bibitem{SoumaBook}
Aoyama, H., Fujiwara, Y., Ikeda, Y., Iyetomi, H., Souma, W., Econophysics and companies: statistical life and death in complex business networks. {\em Cambridge University Press} {\bf 2010}.

 \bibitem{AusloosIvanova03dynamical} Ausloos, M., Ivanova, K., Dynamical model and nonextensive statistical mechanics of a market index on large time windows. {\em Physical Review E} {\bf 2003}, {\em 68}, 046122.

\bibitem{mbraz}
Ausloos, M., Miskiewicz, J., Introducing the q-Theil index. {\em Brazilian Journal of Physics}  {\bf 2009},  {\em 39}(2A), 388-395.

\bibitem{Barbosa}
Barbosa, C. S., Caraballo, R., Alves, L. R., Hartmann, G. A., Beggan, C. D., Viljanen, A., Ngwira, C. M., Papa, A. R. R. Pirjola, R. J., The Tsallis statistical distribution applied to geomagnetically induced currents. {\em Space Weather}  {\bf 2017}, {\em 15}, 1094-1101.

\bibitem{Bebchuk}
Bebchuk, L. A., Kraakman, R., Triantis, G., Stock pyramids, cross-ownership, and dual class equity: the mechanisms and agency costs of separating control from cash-flow rights. In: Concentrated corporate ownership, R.K. Morck (ed.),  {\em University of Chicago Press}, {\bf 2017}, pp. 295-318..

\bibitem{Bellenzier} Bellenzier, L., Grassi R., Interlocking directorates in Italy: persistent links in network dynamics. {\em Journal of Economic Interaction and Coordination} {\bf 2014}, {\em 9}, 183-202.

\bibitem{Bila}
Bila, L., Grech, D., Podhajska, E., Methods of Non-Extensive Statistical Physics in Analysis of Price Returns on Polish Stock Market. {\em Acta Physica Polonica A}  {\bf 2016}, {\em 129}, 986–992.

\bibitem{Caldarelli} Caldarelli,  G.,  Scale-Free Networks: Complex Webs in Nature and Technology, {\em Oxford University Press}, {\bf 2007}

\bibitem{ceptu1}
Ceptureanu, E.G., Ceptureanu S.I., Popescu D., Relationship between Entropy, Corporate Entrepreneurship and Organisational Capabilities
in Romanian Medium Sized Enterprises. {\em Entropy} {\bf 2017}, {\em 19}, 412.

\bibitem{ceptu2}
Ceptureanu, S.I., Ceptureanu, E.G., Marin I., Assessing role of strategic choice on organisational performance by Jacquemin- Berry
entropy index. {\em Entropy} {\bf 2017}, {\em 19}, 448.

\bibitem{Entropy1} Cerqueti, R., Rotundo, G., Ausloos, M.,
Investigating the configurations in cross-shareholding: a joint copula-entropy approach. {\em Entropy} {\bf 2017}, {\em 20}, 134.  

\bibitem{Chang} 
Chang, X., Wang, H., Cross-Shareholdings Structural Characteristic and Evolution Analysis Based on Complex Network. {\em Discrete Dynamics in Nature and Society} {\bf 2017},  5801386. 

\bibitem{Chapelle}  Chapelle, A., Szafarz, A., Controlling Firms Through the Majority Voting Rule. {\em Physica A}  {\bf 2005}, {\em 355}, 509-529.

\bibitem{Cimini} Cimini, G., Serri, M., Entangling credit and funding shocks in interbank markets, {\em PloS One} {\bf 2016},  {\em 11},  e0161642.

\bibitem{cinelli1}
Cinelli, M., Ferraro, G., Iovanella, A., Rich-club ordering and the
dyadic effect: Two interrelated phenomena. {\em Physica A:
Statistical Mechanics and its Applications} {\bf 2018} {\em 490}, 808-818.

\bibitem{cinelli2}
Cinelli, M., Ferraro, G., Iovanella, A., Structural bounds on the dyadic effect. {\em Journal of Complex Networks} {\bf 2017}, {\em 5}, 694-711.

\bibitem{clayton}
Clayton, D.G., A model for association in bivariate life tables and its application in epidemiological studies of familial tendency in
chronic disease incidence. {\em Biometrika} {\bf 1978},  {\em 65}, 141-151.

\bibitem{Clementi} Clementi, F., Gallegati, M., Pareto's law of income distribution: Evidence for Germany, the United Kingdom, and the United States.  In: Chatterjee A., Yarlagadda S., Chakrabarti B.K. (eds)   {\em Econophysics of wealth distributions}, Springer, {\bf 2005},  pp. 3-14.

\bibitem{contreras}
Contreras, M.G.A., Fagiolo, G., Propagation of economic shocks in input-output networks: A cross-country analysis. {\em Physical Review E} {\bf 2014}, {\em 90}, 062812. 

\bibitem{Crane} Crane, D. A.  Has the Obama Justice Department reinvigorated antitrust enforcement?. {\em Stanford  Law Review Online} {\bf 2012}, {\em 65}, 13-20.

\bibitem{Croci} Croci, E., Grassi, R., The economic effect of interlocking directorates in Italy: new evidence using centrality measures. {\em  Computational and Mathematical Organisation Theory}  {\bf 2014},  {\em 20},  89-112.

\bibitem{Serri} D'Arcangelis, A.M., Rotundo, G., Systemic Risk of Non Performing Loans Market. The Italian case. {\em Journal of Applied Quantitative Methods}  {\bf 2019}, {\em  14}(1).  $http://jaqm.ro/issues/volume-14,issue-1/0\_A\_G.PHP$

\bibitem{Derrico}
D'Errico, M., Grassi, R., Stefani, S.,  Torriero, A.,
 Shareholding Networks and Centrality: An Application to the Italian Financial
Market. In: {\em Networks, Topology and Dynamics}.  Lecture Notes in Economics and Mathematical Systems, {\em 613},
A. K. Naimzada, S. Stefani, and  A. Torriero, Eds.,  {\bf 2009}, 215-228.

\bibitem{delpini}
Delpini, D., Battiston, S., Riccaboni, M., Gabbi, G., Pammolli, F., Caldarelli, G.,
Evolution of controllability in interbank
networks. {\em Scientific Reports}  {\bf 2013},  {\em 3}, 1626.

 \bibitem{Devi19} Devi, S., Financial portfolios based on Tsallis relative entropy as the risk measure. {\em Journal of Statistical Mechanics: Theory and Experiment} {\bf 2019},   093207.

\bibitem{Elliott} Elliott,  M., Golub, B., Jackson, M.O., Financial networks and contagion.  {\em The American Economic Review} {\bf 2014}, {\em 104},  3115-3153.

\bibitem{Feller} Feller, W., {\em An Introduction to Probability Theory and its Applications II}  (2nd ed.). New York: Wiley. {\bf 1971}

\bibitem{ferraro2017}
Ferraro, G., Iovanella, A., Technology transfer in innovation networks: An empirical study of the Enterprise Europe Network. {\em
International Journal of Engineering Business Management} {\bf 2017}, {\em 9}, 1-14. 

\bibitem{ferraro2015}
Ferraro, G., Iovanella, A., 
Organizing collaboration in inter-organisational innovation networks, from orchestration to choreography. 
{\em International Journal of Engineering Business Management} {\bf 2015}, {\em 7}, 7-24.

\bibitem{frank}
Frank, M.J., On the simultaneous associativity of $F(x, y)$ and $x+y- F(x,y)$. {\em Aequationes Math.}  {\bf 1979}, {\em 19},
194-226.

\bibitem{frechet}
Frechet, M.,  Remarques au sujet de la note pr\'ec\'edente. {\em C.R. Acad. Sci. Paris} {\bf 1958}, {\em 246}, 2719-2720.

\bibitem{Axel} Gandy, A., Veraart, L.A.M., A Bayesian methodology for systemic risk assessment in financial networks. 
{\em Management Science} {\bf 2016}, {\em 63}, 4428-4446.

\bibitem{gao}
Gao, J., Barzel, B., Barabasi, A.L., Universal resilience patterns in complex networks. {\em Nature}  {\bf 2016}, {\em 530}
(7590), 307-312.

\bibitem{Garlaschelli}
Garlaschelli, D., Battiston, S., Castri, M., Servedio, V., Caldarelli, G.,  The scale-free topology of market investments. {\em Physica A: Statistical Mechanics and its Applications} {\bf 2005},  {\em 350}, 491-499.


\bibitem{Gell-Mann}
Gell-Mann, M., Tsallis, C. (Eds.), Nonextensive entropy: interdisciplinary applications. {\em Oxford University Press on Demand}, {\bf 2004}.

\bibitem{Gulati} Gulati, R., Westphal, J.D., Cooperative or controlling? The effects of CEO-board relations and the content of interlocks on the formation of joint ventures, {\em Administrative Science Quarterly} {\bf 1999}, {\em  44}, 473-506.

\bibitem{gumbel}
Gumbel, E.J.. Bivariate exponential distributions. {\em J. Amer. Statist. Assoc.} {\bf 1960}, {\em 55}, 698-707.

\bibitem{iori}
Iori, G., De Masi, G., Precup, O.V., Gabbi, G.,  Caldarelli, G.,
A network analysis of the Italian overnight money market. {\em Journal of Economic Dynamics and Control} {\bf 2008}, {\em 32}, 259-278.

\bibitem{Jakimowicz} Jakimowicz, A., The Role of Entropy in the Development of Economics. {\em Entropy}  {\bf 2020}, {\em 22},  452.

\bibitem{joe}
Joe, H., Multivariate models and multivariate dependence concepts.
{\em CRC Press},  {\bf 1997}.

\bibitem{Li}
    Li, H., Fang, W., An, H., Yan, L., The shareholding similarity of the shareholders of the worldwide listed energy companies based on a two-mode primitive network and a one-mode derivative holding-based network. {\em Physica A: Statistical Mechanics and its Applications} {\bf 2014},  {\em 415}, 525-532.

    \bibitem{LiAn}
Li, H., An, H., Gao, X., Huang, J., Xu, Q., On the topological
properties of the cross-shareholding networks of listed companies in China: Taking shareholders' cross-shareholding relationships into
account.  {\em Physica A: Statistical Mechanics and its Applications}  {\bf  2014}, {\em 406}, 80-88.


\bibitem{Lichtenberg}
 Lichtenberg, F. R.,  Pushner, G. M., Ownership structure and corporate performance in Japan. {\em  Japan and the World Economy}  {\bf 1994},  {\em 6}, 239-261.

\bibitem{liese}
Liese, F., Vajda, I., Convex Statistical Distances. {\em B.G. Teubner Verlagsgesellschaft, Leipzig},  {\bf 1967}.

\bibitem{ling}
Ling, C.H., Representation of associative functions.  {\em Publ. Math. Debrecen}  {\bf 1965}, {\em 12}, 189-212.

\bibitem{luo}
Luo, J., The power‐of‐pull of economic sectors: A complex network analysis. {\em Complexity} {\bf 2013}, {\em 18}, 37-47.

   \bibitem{Ma}
Ma, Y., Zhuang, X.,  Li, L.,  Research on the relationships of the
domestic mutual investment of China based on the cross-shareholding networks of the listed companies. {\em Physica A: Statistical
Mechanics and its Applications}  {\bf 2011}, {\em 390}, 749-759.

 \bibitem{MaasoumiRacine02} Maasoumi, E., Racine, J.,  Entropy and predictability of stock market returns. {\em Journal of Econometrics} {\bf  2002}, {\em   107}, 291-312.

\bibitem{pone1}
Maluck, J., Donner, R.V., A network of networks perspective on global trade. {\em PloS One}  {\bf 2015}, {\em 10},  e0133310.

\bibitem{Masz}
Maszczyk, T., Duch, W., Comparison of Shannon, Renyi and Tsallis Entropy Used in Decision Trees. In: Rutkowski L., Tadeusiewicz R., Zadeh L.A., Zurada J.M. (eds) {\em Artificial Intelligence and Soft Computing – ICAISC 2008}. Lecture Notes in Computer Science {\em 5097}. Springer, Berlin, Heidelberg, {\bf 2008}, pp. 643-651.

\bibitem{nelsen}
Nelsen, R.B., An Introduction to Copulas. {\em Springer New York}, {\bf 1999}.

\bibitem{Newman} Newman, M., Barabasi, A.L., Watts, D.J., The structure and dynamics of networks, {\em Princeton University Press}, {\bf 2011}.

\bibitem{ohn}
Ohnishi, T., Takayasu, H., Takayasu, M., Hubs and authorities on Japanese inter-firm network: Characterisation of nodes in very large directed networks. {\em Progress of Theoretical Physics Supplement} {\bf 2009}, {\em 179}, 157-166.

\bibitem{Okabe}
 Okabe, M., Cross Shareholdings in Japan. A New Unified Perspective of the Economic System, {\em Edward Elgar Publishing}, {\bf 2002}.

\bibitem{Pastor} Pastor-Satorras, R., Vespignani, A., Epidemic dynamics and endemic states in complex networks. {\em Physical Review E} {\bf 2001}, {\em 63}, 066117.

\bibitem{Pastor1} Pastor-Satorras, R., Vespignani, A., Epidemic spreading in scale-free networks. {\em Physical Review Letters} {\bf 2001}, {\em  86},  3200.

\bibitem{Pavlos}
 Pavlos, G. P., Karakatsanis, L. P., Iliopoulos, A. C., Pavlos, E. G.,  Tsonis, A. A., Non-extensive statistical mechanics: Overview of theory and applications in seismogenesis, climate, and space plasma. In Tsonis A. (ed.), {\em Advances in Nonlinear Geosciences} {\bf 2018}, {\em 465-495}. Springer, Cham.
 
 \bibitem{Petroni}
 Petroni, F., Ausloos, M., High frequency (daily) data analysis of the Southern
 Oscillation Index. Tsallis nonextensive statistical mechanics approach. {\em The European Physical Journal Special Topics} {\bf 2007}, {\em 143}, 201-208.
 
 \bibitem{Queiros}
  Queiros, S. D., Moyano, L. G., De Souza, J.,  Tsallis, C., A nonextensive approach to the dynamics of financial observables. {\em The European Physical Journal B}  {\bf 2007}, {\em 55}, 161-167.
  
  \bibitem{Rak2007}
    Rak, R., Dro{\.z}d{\.z}, S.,   Kwapie{\'n}, J., Nonextensive statistical features of the Polish stock market fluctuations. {\em Physica A: statistical mechanics and its applications}  {\bf 2007}, {\em  374}, 315-324.
  
  \bibitem{Rak2013}
    Rak, R., Dro{\.z}d{\.z}, S., Kwapie{\'n}, J., Oswiecimka, P., Stock returns versus trading volume: is the correspondence more general?. {\em Acta Phys. Pol. B}  {\bf 2013}, {\em  44},  2035-2050.
    
\bibitem{rachev}
Rachev, S.T., Probability Matrices and the stability of stochastic models. {\em Wiley}, {\bf 1991}.

 \bibitem{EPJB86.13GRMAcomplex}
 Rotundo, G.,   Ausloos, M., Complex-valued information entropy measure for networks with directed links (digraphs). 
 Application to citations by community agents with opposite opinions, {\em European Physical  Journal   B} {\bf 2013}, {\em 86},  169.

\bibitem{GRAMD} 
Rotundo, G., D'Arcangelis, A.M., Ownership and control in shareholding networks. 
{\em Journal of Economic Interaction and Coordination}  {\bf 2010}, {\em 5},   191-219.

\bibitem{GRAMDAASS} Rotundo, G., D'Arcangelis, A.M.,  Network analysis of ownership and control structure in the
Italian Stock market.  {\em Advances and Applications in Statistical Sciences} {\bf 2010}, {\em  2},  255-273.

\bibitem{GRAMDQuQu} Rotundo, G., D'Arcangelis, A.M., Network of companies: an analysis of market concentration in the
Italian stock market. {\em  Quality \& Quantity} {\bf 2014}, {\em 48}, 1893-1910.

\bibitem{Ruseckas}
  Ruseckas, J., Gontis, V.,   Kaulakys, B., Nonextensive statistical mechanics distributions and dynamics of financial observables from the nonlinear stochastic differential equations. {\em Advances in Complex Systems}  {\bf 2012}, {\em 15}(S01), 1250073.

\bibitem{schellcase} Schellhase, C., Density and copula estimation using penalised spline smoothing, {\bf 2012}. Available at: $pub.uni-bielefeld.de$. 

\bibitem{Shannon} Shannon, C. E., Weaver, W., The Mathematical Theory of Communication. {\em Urbana, University of Illinois Press},  {\bf 1949}.

\bibitem{silva1} Silva, T.C., Alexandre, M.D.S., Tabak, B.M., Bank lending and systemic risk: A financial-real sector network approach with feedback.  {\em Journal of Financial Stability} {\bf 2018}, {\em 38}, 98-118. 

\bibitem{sklar}
Sklar, M., Fonctions de r{\'e}partition \`a $n$ dimensions et leurs marges.
{\em Publ. Inst. Statist. Univ. Paris} {\bf 1959}, {\em 8}, 229-231.

\bibitem{soromaki}
Soramaki, K., Bech, M.L., Arnold, J., Glass, R.J., Beyeler, W.E.,
The topology of interbank payment flows. {\em Physica A: Statistical Mechanics and its Applications} {\bf 2007}, {\em 379}, 317-333.

\bibitem{Souma}
Souma, W., Fujiwara, Y., Aoyama H.,  Shareholding Networks in Japan.  {\em AIP Conference Proceedings} {\bf 2005}, {\em 776},  298-307, 

\bibitem{Soumab}
Souma, W., Fujiwara, Y., Aoyama, H., Change of ownership networks in Japan.   In: Takayasu H. (eds)  {\em Practical Fruits of Econophysics}  {\bf 2006}, 307-311

\bibitem{silva2}
Souza, S.R.S.D., Silva, T.C., Tabak, B.M., Guerra, S.M.,
 Evaluating systemic risk using bank default probabilities in financial networks.
{\em Journal of Economic Dynamics and Control} {\bf 2016}, {\em 66}, 54-75.

\bibitem{Tsallis}
Tsallis, C., Possible generalisation of Boltzmann-Gibbs Statistics.  {\em Journal of Statistical Physics}  {\bf 1988}, {\em 52},  479-487.

\bibitem{Tsallis1}
Tsallis, C., Economics and Finance: q-Statistical Stylized Features Galore. {\em  Entropy}  {\bf 2017}, {\em 19},  457.

\bibitem{US}  US mergers     guidelines.
    $http://www.stanfordlawreview.org/online/obama-antitrust-enforcement$.
    
 \bibitem{Vinte}   Vințe, C.,  Smeureanu, I.,  Furtună, T.-F., Ausloos, M., An Intrinsic Entropy Model for Exchange-Traded Securities. {\em Entropy} {\bf  2019}, {\em 21}, 1173. 

\bibitem{Vitali}
 Vitali, S., Glattfelder, J.B., Battiston, S.,
The network of global corporate control. {\em PLoS ONE} {\bf 2011}, {\em 6}, e25995. 

\bibitem{weber}
Weber, S., Weske, K., The joint impact of bankruptcy costs, fire sales and cross-holdings on systemic risk in financial networks.
{\em Probability Uncertainty and Quantitative Risk}  {\bf 2017}, {\em 2}, 9. 

\bibitem{complexq}
Wilk, G.,    Włodarczyk, Z.,
Tsallis distribution with complex nonextensivity parameter q.
{\em Physica A: Statistical Mechanics and its Applications} {\bf 2014}, {\em 413}, 53-58.

\bibitem{zambrano} Zambrano, E., Hernando, A., Fern\'andez Bariviera, A., Hernando, R., Plastino, A.,
 Thermodynamics of firms' growth. {\em J. R. Soc. Interface} {\bf 2015}, {\em 12}, 20150789.

 \bibitem{zhou13applications} Zhou, R., Cai, R., Tong, G., Applications of entropy in finance: A review. {\em Entropy} {\bf 2013}, {\em 15}, 4909-4931.

\end{thebibliography}


\end{document}